\documentclass[11pt]{article}

\usepackage{a4wide}
\usepackage{amsmath,amssymb,amsthm,amscd}
\usepackage{graphicx}
\usepackage{authblk}

\usepackage{natbib}
\usepackage{algorithm}	% those two packages are helpful if you use a lot of algorithms in your thesis
\usepackage{algorithmic}
\usepackage{bm}
\usepackage{framed}
\usepackage{hyperref}

\usepackage[toc,page]{appendix}

\graphicspath{ {Images/} }

\theoremstyle{plain}

\begin{document}

\title{Approximate methods for dynamic ecological models}

\author{Matteo Fasiolo}
\author{Simon N. Wood}
\affil{Department of Mathematical Sciences, University of Bath}
\affil{Correspondence: matteo.fasiolo@bath.edu}

\maketitle

\abstract{
This document is due to appear as a chapter of the forthcoming Handbook of Approximate Bayesian Computation (ABC) by S. Sisson, L. Fan, and M. Beaumont. Here we describe some of the circumstances under which statistical ecologists might benefit from using methods that base statistical inference on a set of summary statistics, rather than on the full data. We focus particularly on one such approach, Synthetic Likelihood, and we show how this method represents an alternative to particle filters, for the purpose of fitting State Space Models of ecological interest. As an example application, we consider the prey-predator model of \cite{turchin2000}, and we use it to analyse the observed population dynamics of Fennoscandian voles\footnotemark[1]. 
}


\footnotetext[1]{The data on voles abundance and the code for this model can be found at \url{https://github.com/mfasiolo/volesModel}.}

\section{Simulation-based methods in ecology}

\subsection{Intractable ecological models}

Ecology aims to understand the abundance and distribution of organisms \index{ecology}. This essentially quantitative task is made difficult by the complex web of interactions that exist between living things. In the face of such daunting ecological complexity, dynamic models play an important role in separating fundamental mechanisms from matters of detail. In particular, they allow theoretical ideas to be sharpened into well defined quantitative hypotheses, and this in turn opens up the possibility of testing these hypotheses using data. 

But there is a catch. To be useful, ecological dynamic models must often resort to `cartooning' of some ecological processes. Simplification is essential if the model is not to become a `model-of-everything', hence a reasonably parsimonious model may not be intended to reproduce the full data $y_{obs}$ in all its features. For example, while the full data might be characterized by a spatial or temporal structure, it is often convenient to use a lumped model that ignores these dimensions. Similarly, when the data contains several classes of organisms, computational considerations might lead to a model that aggregates key statistics, such as population counts, over different classes. Under these circumstances, reducing the full data to a set of summary statistics, \index{summary statistics} $s_{obs} = S(y_{obs})$, might not lead to any loss of information during parameter estimation or model selection \cite{hartig2011statistical}.

Basing statistical inference on aggregate summary statistics might be necessary even when working with individual based models, \index{individual based models} which are often used to understand ecological outcomes that depend intricately on the interactions of individuals within a population. Forest stand growth models are an example. In these models individual trees of many species may be grown to maturity, all competing continuously for light and nutrients as they do so. Here the mismatch between data and model is of a different kind. For example, in a real forest we would obtain data consisting of measurements on individual trees. The same measurements can often be made on the model trees, but a particular model individual does not correspond to any real individual. We are left with no choice but to base inference on summary statistics, which suggests the use of ABC-type methods. One such example is \cite{hartig2014technical} who uses Synthetic Likelihood \citep{wood2010statistical}, an approximate method closely related to ABC, to fit the Formind individual-based forest model to Ecuadorian tropical forest field data. While the model deals with individual trees, its output is summarized using 112 statistics such as biomass, growth rate and tree counts, obtained by aggregating trees over several diameter classes. 

Other reasons for considering the use of summary statistics relate to highly non-linear dynamics, \index{dynamics!non-linear} of the sort that are often found in populations of small animals, with high rates of fecundity and mortality. Indeed, even if our models are perfect descriptions of the driving ecological mechanisms, dynamic irregularity can make reliable inference very difficult to achieve by conventional means. If our models are less perfect, the interaction of such irregularity with small infelicities in the model's ability to match the data can lead to substantial inferential errors. 
\cite{wood2010statistical} shows that these problems can arise in ecological systems as simple as the Ricker map \citep{may1976simple}, and illustrates how the extreme sensitivity of near chaotic systems to small changes in dynamically important parameters can cause minuscule moves in the parameter space to result in massive changes in likelihood values. In this circumstance, it is obviously appealing to base inference on summary statistics of the data that the model should be able to reproduce, rather than on the full data. Indeed, \cite{wood2010statistical} and \cite{fasiolo2014statistical} argue that ABC-type methods can offer an appealing robustness here, provided that they are used in conjuction with appropriately robust statistics. \index{summary statistics!robustness}

Even in the absence of the difficuties just discussed, ecological models can have tractability problems. Most of the conventional statistical tools used to find the parameter values or models that are most consistent with data (and possibly with prior knowledge), rely on the likelihood function, $p(y_{obs}|\theta)$. Unfortunately, for many models of ecological interest, $p(y_{obs}|\theta)$ is not available directly or is otherwise problematic, thus posing an obstacle to the whole inferential process. This difficulty can occur for several possible reasons, but one common problem is the presence of hidden or latent states. Specifically, we often know that the dynamics of an observed process $y_{obs}$ are related to those of other processes $n$, which are hidden from us. In such cases the likelihood could ideally be obtained by integrating the latent states out of the joint probability density of data and hidden states \index{hidden states}
\begin{equation}
p(y_{obs}|\theta) = \int p(y_{obs}, n|\theta)\, d n.
\end{equation}
In practice this integration problem is usually analytically intractable, while the efficient implementation of numerical or Monte Carlo integration schemes often require additional assumptions, such as those detailed in Section \ref{sec:SSMS}.

\begin{minipage}{0.95\linewidth}
\begin{framed} \label{ABCreject}
\begin{center}
\emph{Box 1.1: An ABC rejection sampler}
\end{center} 
The most basic instance of an ABC sampler, targeting an approximation to $p(\theta|s_{obs})$, is the following rejection algorithm:
\begin{enumerate}

\item Sample $M$ parameter vectors $\theta^1, \dots, \theta^M$ from the prior $\pi(\theta)$.

\item For each parameter vector $\theta^i$, with $i = 1,\dots,M$, simulate a corresponding datasets $Y^i$ from $p(y|\theta^i)$.

\item Transform the simulated datasets $Y^1, \dots, Y^M$ to vectors of summary statistics $S^1=S(Y^1), \dots, S^M=S(Y^M)$. 

\item Calculate the distances $d^i = d(s_{obs}, S^i)$, for $i = 1, \dots, M$, using an appropriate distance measure $d(\cdot, \cdot)$. 

\item Store the $N \leq M$ parameter vectors $\theta^{i_1},\dots,\theta^{i_N}$, whose corresponding distances $d^{i_1}, \dots, d^{i_N}$ are all lower than a tolerance $h>0$.
\end{enumerate}
Notice that $\theta^{i_1},\dots,\theta^{i_N}$ are effectively a sample from  
$$
\pi_{ABC}(\theta|s_{obs}) \propto p\{d(s_{obs}, s)<h|\theta\} \pi(\theta),
$$ 
which should be a close approximation to $\pi(\theta|s_{obs})$, if $h$ is sufficiently small.
\end{framed}
\end{minipage}  

\vspace{10pt}
Classical examples of partially observed systems of ecological interest are predator-prey systems, where the abundance of one of the two components is often completely unknown. In Section \ref{sec:voles} we consider the prey-predator model proposed by \cite{turchin2000}, which has been used to describe the population dynamics \index{dynamics!population} of Fennoscandian voles. In that example trap data provides noisy estimates of voles abundance, but no such proxy is available for predatory weasels. A similar example is provided by \cite{kendall2005population}, who evaluate alternative explanations for the regular oscillations in population density of insect pest pine looper moths. They consider, among others, a parasitoid and a food quality model and they fit them using only data on moth population density. Given that ecological systems are observed with noise in most cases, the issue of hidden states is widespread and it appears in studies concerned with animal movement \citep{langrock2012flexible,morales2004extracting}, population abundance estimation \citep{farnsworth2007estimate}, and essentially whenever remote tracking data is available \citep{jonsen2005robust}.

The rapid growth in computational resources has supported the development of several approaches meant to tackle the issue of intractable likelihoods. Some of these approaches exploit the fact that faster computation makes forward model simulation, that is simulation of data $y$ from $p(y|\theta)$, cheap enough that it can be repeated many thousands of times. In particular, it is possible to use forward simulations to find the set of parameter values or models that are able to closely reproduce the full data, $y_{obs}$, or more often some of its most informative features, $s_{obs}$. ABC represents one class of such methods which, being based on a Bayesian framework, generally try to address questions regarding parameter estimation or model selection by approximately sampling the corresponding posteriors $p(\theta|s_{obs})$ and $p(Mod|s_{obs})$. The rejection sampler described in Box \ref{ABCreject} is probably the simplest exponent of the ABC family.

%Indeed, matching the full data $y_{obs}$ via ABC is seldom feasible, and as we have discussed, in many ecological settings there will anyway be good reasons to focus on carefully chosen summary statistics $s_{obs}$ instead.  In a nutshell this is the solution offered by ABC methods, whose simplest instance is the ABC rejection sampler described in Box \ref{ABCreject}, to parameter estimation and model selection problems.

In the remainder of this chapter we focus on a particular family of intractable models: state space models. In Section \ref{sec:SSMS} we briefly describe this class of partially observed models, which are very popular in the ecological literature, and we introduce two approaches that can be used to perform statistical inference for such models. In Section \ref{sec:SLvsABC} we discuss how one of these approaches, synthetic likelihood (SL), differs from other ABC methods, while in Section \ref{sec:voles} we consider the predator-prey model of \cite{turchin2000} and compare the available methods using both simulated and field data. Finally, in Section \ref{ref:discussion}, we conclude by making some practical consideration regarding the benefits and drawbacks of using ABC or SL, rather than less approximate methods, when working with state space Models. 

\subsection{Inference for state space models} \label{sec:SSMS}

State space models \index{state space models} (SSMs) represents a special class of models with hidden or partially observed states. In these models the hidden states follow Markov processes, whose conditional pdf has the following property
\begin{equation} \label{eq: markov_prop}
p(n_{t} | n_{1}, \dots, n_{t-1}, \theta) = p(n_{t} | n_{t-1}, \theta),
\end{equation}
where $t \in \{1, \dots, T\}$ and $\theta$ is a vector of static parameters. Property (\ref{eq: markov_prop}) implies that the future states are statistically independent of the past, upon conditioning on the present. Generally, the hidden ecological processes are coupled with an observation process according to which observed data points are conditionally independent, given the underlying states \citep{king_annual}
\begin{equation}
p(y_{t} | n_t, y_{1}, \dots, y_{t-1}, \theta) =  p(y_{t} | n_t, \theta),
\end{equation}
where we defined $y_{t} = y_{obs, t}$, to simplify the notation. Typically the term SSMs is used to indicate partially observed Markov processes with continuous state spaces, while models with discretely valued states are called hidden Markov models \index{hidden Markov models} (HMMs). In the following we focus on SSMs, but most considerations apply also to HMMs.

As for most partially observed systems, the likelihood of SSMs is generally not available directly. Indeed for such models $p(y_{1:T} | \theta)$, where $y_{1:T} = \{y_1, \dots, y_T\}$, is available analytically only if both $p(n_{t} | n_{t-1}, \theta)$ and $p(y_{t} | n_t, \theta)$ are linear and Gaussian \citep{kalman1960new}. Fortunately, the Markov property (\ref{eq: markov_prop}) mitigates the intractability of these models, because it allows  estimation of the likelihood by performing the required $T$-dimensional integration efficiently. In particular, the Markov property is exploited by particle filters \index{particle!filters} \citep{doucet2009tutorial} to break down the integration problem into $T$ sequential integration steps. These computational tools can be used to obtain Monte Carlo estimates $\hat{p}(y_{1:T}|\theta)$ of the full likelihood function. We describe the Sequential Importance Re-Sampling (SIR) algorithm, \index{sequential importance re-sampling} which is the simplest instance of a particle filter, in Box \ref{SIRalgo}.

A more general solution to the problem of intractable likelihoods is offered by SL \citep{wood2010statistical}. \index{SL!introduction} This is a simulation-based and approximate approach, which is closely related to ABC methods. 
Rather than approximating the full likelihood function, SL transforms the data to a set of summary statistics $s_{obs}$, and approximates $p(s_{obs}|\theta)$ parametrically. In particular, SL assumes that the summary statistics are approximately normally distributed, conditionally on the parameters
\begin{equation} \label{eq:normApprox}
S \sim \text{N}\big \{ \mu (\theta), \Sigma (\theta) \big \},
\end{equation} 
where the functions $\mu(\theta)$ and $\Sigma(\theta)$ are generally unknown. Given that the parametric density assumption does not hold exactly in general, the resulting synthetic likelihood, $p_{SL}(s_{obs}|\theta)$, should be considered an approximation to $p(s_{obs}|\theta)$. Point-wise estimates of the synthetic likelihood can be obtained by using the procedure described in Box \ref{SLalgo}.

\begin{minipage}{0.95 \linewidth}
\begin{framed} \label{SIRalgo}
\begin{center}
\emph{Box 1.2: Sequential Importance Re-Sampling (SIR)}
\end{center} 
This algorithm was proposed by \cite{gordon1993novel}, and has been hugely successful in the context of SSMs. It implements a sequential importance sampling procedure, with a re-sampling step that is used to discard particles with low weights, thus mitigating the particle depletion problem \citep{doucet2009tutorial}. An estimate of the likelihood at $\theta$ can be obtained using the following steps
\begin{enumerate}
\item Draw particles $N_0^i$, for $i = 1, \dots , M$, from the prior distribution of the initial state $N_0^i \sim \pi (n_0)$.
\item For $t = 1$ to T:
\begin{enumerate}
\item \emph{Prediction step}: propagate the particles forward in time 
$$
N_t^i \sim p(n_t|n_{t-1}^i, \theta),\;\;\; \text{for} \;\;\; i = 1, \dots, M.
$$ 
\item \emph{Update step}: calculate the normalized weight of each particle
$$
w^i = \frac{ \tilde{w}^i }{ \sum_{i = 1}^N \tilde{w}^i }, \;\;\; \tilde{w}^i = p(y_t|n_t^i,\theta),\;\;\; \text{for} \;\;\; i = 1, \dots, M.
$$
\item Estimate the current component of the likelihood
$$
\hat{p}(y_t|y_{1:t-1}, \theta) = \frac{1}{M} \sum_{i = 1}^M \tilde{w}^i. 
$$
\item Re-sample the particles multinomially with replacement, using probabilities equal to the normalized weights.
\end{enumerate} 
\item Estimate the likelihood using the decomposition
$$
\hat{p}(y_{1:T}|\theta) = \hat{p}(y_1|\theta) \prod_{t = 2} ^ T \hat{p}(y_t|y_{1:t-1}, \theta).
$$
\end{enumerate}  
\end{framed}
\end{minipage}
\begin{minipage}{0.95 \linewidth}
\begin{framed} \label{SLalgo}
\begin{center}
\emph{Box 1.3: Evaluating the synthetic likelihood}
\end{center} 
Point-wise estimates of the synthetic likelihood, at an arbitrary position $\theta_p$ in the parameter space, can be obtained as follows:
\begin{enumerate}
\item simulate $M$ datasets $Y^1,\dots,Y^M$ from the model $p(y|\theta_p)$ and transform them into $d$-dimensional summary statistic vectors $S^1=S(Y^1),\dots,S^M=(Y^M)$.
\item Estimate mean and covariance matrix of the summary statistics, using standard estimators
$$
\hat{\mu}(\theta_p) = \frac{1}{M} \sum_{i=1}^M S^i,
$$
$$
\hat{\Sigma}(\theta_p) = \frac{1}{M-1} \sum_{i=1}^M \big \{ S^i - \hat{\mu}(\theta_p) \big \} \big \{ S^i - \hat{\mu}(\theta_p) \big \}^T,
$$
or possibly more robust alternatives.
\item Evaluate the corresponding Gaussian density at the observed statistics, that is
\begin{equation*}
\begin{split}
\hat{p}_{SL}(s_{obs}|\theta_p) &= (2\pi)^{-\frac{d}{2}}|\hat{\Sigma}(\theta_p)|^{-\frac{1}{d}}\\ &\times \exp{\bigg [-\frac{1}{2}\big\{s_{obs}-\hat{\mu}(\theta_p)\big\}^T\hat{\Sigma}(\theta_p)^{-1}\big\{s_{obs}-\hat{\mu}(\theta_p)\big\}\bigg ]}.
\end{split}
\end{equation*}
\end{enumerate}
\end{framed}
\end{minipage}
\vspace{10pt}

There exists a strong relationship between SL and the simulation-based approach of \cite{diggle1984monte}, who proposed to estimate the full likelihood $p(y_{obs}|\theta)$ point-wisely, by simulating data from the model and approximating its distribution using a non-parametric density estimator. Most ABC algorithms follow a less likelihood-centric approach, because they generally aim at sampling from $\pi(\theta|s_{obs})$ directly. This is the case, for instance, in ABC rejection, Markov Chain Monte Carlo (MCMC) and Sequential Monte Carlo (SMC) algorithms \citep{beaumont2010approximate}. Section \ref{sec:SLvsABC} discusses how SL differ from other ABC methods in more details.

The point estimates $\hat{p}(y_{obs}|\theta)$ and $\hat{p}_{SL}(s_{obs}|\theta)$, obtained using respectively SIR and SL, can be used within a Metropolis Hastings (MH) algorithm. Specifically, if SL is used, the MH acceptance probability is given by
\begin{equation} \label{eq:acceptRatio}
\alpha = \text{min} \bigg \{ 1, \frac{ \hat{p}_{SL}(s_{obs}|\theta^*) p(\theta|\theta^*) \pi (\theta^*)}{ \hat{p}_{SL}(s_{obs}|\theta)p(\theta^*|\theta) \pi (\theta) } \bigg \},
\end{equation}
where $p(\theta^*|\theta)$ is the transition kernel and $\pi (\theta)$ is the prior density. When $\hat{p}(y_{obs}|\theta)$ is used in place of $\hat{p}_{SL}(s_{obs}|\theta)$ in (\ref{eq:acceptRatio}), the resulting sampler is called a particle marginal Metropolis Hastings \index{particle!PMMH} (PMMH) algorithm \citep{andrieu2010particle}. Under the assumptions detailed by \cite{andrieu2009pseudo}, this sampler targets $\pi(\theta|y_{obs})$, thus representing an exact-approximate algorithm. When SL is used the situation is more complex because, unless the statistics are normally distributed across the parameter space, the resulting synthetic likelihood Metropolis Hasting (SLMH) \index{SL!SLMH} algorithm will target $\pi(\theta|s_{obs})$ only approximately.

The main drawback of using SLMH or PMMH is their high computational cost: the value of the (synthetic) likelihood function at the proposed parameters $\theta^*$ has to be estimated at each MH step, and this can be expensive for complex models. For this reason \cite{wilkinson2014accelerating} and \cite{gutmann2015bayesian} avoid using SLMH, by explicitly approximating the synthetic likelihood function $\hat{p}(s_{obs}|\theta)$ using Gaussian Processes. Their approaches clearly extend to situations where the likelihood is estimated using a particle filter. An additional complication of MH algorithms using noisy likelihood estimates is that they are often affected by poor mixing, because the sampler tends to get trapped when an unusually high estimate of the likelihood is reached (an ad hoc solution is to simply re-estimate the value of the (synthetic) likelihood at latest accepted position, $\theta$, at every MH step). This problem is discussed by \cite{doucet2015efficient} and \cite{sherlock2014efficiency}, who study how to tune MH algorithms which make use of noisy and unbiased likelihood estimates.  

Given the above issues, ABC methods might appear to be more efficient than SLMH or PMMH, because at each iteration they typically simulate only a single summary statistics vector from $p(s|\theta)$. However, the accuracy and the acceptance ratio of ABC samplers are, respectively, inversely and directly proportional to the tolerance $h$. This trade-off makes it is difficult to formulate a clear statement about the computational efficiency of ABC methods, relative to SLMH and PMMH.

While in \ref{sec:SLvsABC} we discuss the merits and drawbacks of SL relative to other ABC methods, we come back to SLMH and PMMH in Section \ref{sec:voles}, where we use them to fit the SSM of \cite{turchin2000} to ecological data.

\subsection{SL versus tolerance-based ABC} \label{sec:SLvsABC}

\index{SL! vs ABC}

The choice of summary statistics is crucial for the performance of ABC methods, hence the topic has been the subject of much research. See \cite{blum2013comparative} for a comprehensive review of methods for dimension reduction or statistics selection. SL and ABC methods share some requirements regarding the choice of summary statistics. More specifically, in parameter estimation problems the summary statistics should contain as much information as possible about the parameters, so that $\pi(\theta | y_{obs})$ will be approximately proportional to $\pi(\theta| s_{obs})$. 

Beside this common ground, SL differs from ABC methods in several ways, and this entails some diverging requirements on the summary statistics. In particular, reducing the number summary statistics is more critical to ABC methods than to SL. In fact, the non-parametric approach followed by most ABC methods, implies that the convergence rate of the resulting posterior distributions slows down rapidly as the dimension of the statistics vector increases \citep{blum2010approximate}. On the other hand, the parametric likelihood estimator used by SL, ensures that this method is much less sensitive to the number of summary statistics used. This difference in scalability has important practical implications. In particular, SL allows practitioners to focus on the challenging task of identifying informative summary statistics, without having to worry too much about keeping their number low. Obviously SL's scalability in the number of statistics does not come without a cost, but it has to be paid for in parametric assumptions, whose effect might be hard to quantify.  

\index{ABC!tuning}

Another potential issue with ABC algorithms, such as the rejection sampler in Box \ref{ABCreject}, is that they often measure the distance between the observed and simulated statistics using a squared Mahalanobis distance
$$
d(s_{obs}, S) = || s_{obs},  S ||_A^2 = ( s_{obs} -  S)^T  A ( s_{obs} -  S), 
$$
where $ A$ is a scaling matrix.
%, and accept a proposed set of parameters $\theta^*$ if
%$$
%|| s_{obs},  S ||_A^2 < h, \;\;\;\; S \sim p(s|\theta),
%$$
%where $h > 0$ is an acceptance threshold or tolerance. 
The choice of $ A$ is fundamental when the summary statistics have very different scales or when there are subsets of highly correlated statistics. A possible solution is to simulate $N$ vectors of summary statistic at some location $ \theta_p$ in the parameters space and use the inverse of the empirical covariance matrix of the simulated summary statistics as scaling matrix $A = \hat{ \Sigma}(\theta_p)^{-1}$.
This simple choice works well in many cases, but it can lead to unsatisfactory results when the covariance of the summary statistics varies strongly with model parameters.

As an illustration of this problem, we consider a stochastic version of the Ricker map \index{Ricker map}
$$
Y_t \sim \text{Pois}(\phi X_t), \;\;\; N_t = r N_{t-1} e ^ {-N_{t-1} + Z_t}, 
\;\;\; Z_t \sim N(0, \sigma^2),
$$
where $N_t$ is the population size at time $t$, $r$ is the intrinsic growth rate of the population, $\phi$ is a scaling parameter and $Z_t$ can be interpreted as environmental noise. In the following we employ the set of 13 summary statistics proposed by \cite{wood2010statistical}, who used them to fit this model with SL. 

\begin{figure}
\centering
\includegraphics[scale = 0.4, angle=-90]{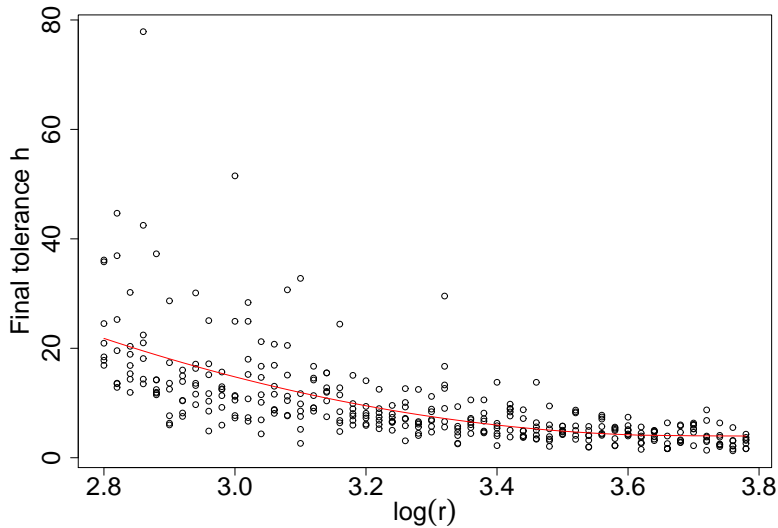}
\caption{Lowest achievable tolerance $h$ versus value of $\log r$ at which the scaling matrix is estimated. The red line is a quadratic regression fit.}
\label{fig: tol_vs_par}
\end{figure}

In order to quantify the importance of the scaling matrix $ A$ in this setting, we performed the following simulation experiment
\begin{itemize}
\item Define a sequence of equally space values $v_k$, for $k = 1, 2, \dots, 50$, ranging from $2.8$ to $3.8$. 
\item For each value $v_k$:
\begin{enumerate}
\item Simulate a path $ Y_{1:T}$ from the Ricker map, using $T = 50$ and parameter values $\log r = 3.8$, $\sigma^2 = 0.3$ and $\phi = 10$. Define $s_{obs} = S(Y_{1:T})$.
\item Set the initial parameter vector $ \theta_p$ to $\log{r} = v_k$, $\sigma^2 = 0.3$ and $\phi = 10$.  
\item Simulate $10^4$ paths from the model using parameters $ \theta_p$, transform each of them into a vector summary statistics and calculate their empirical covariance $\hat{ \Sigma}(\theta_p)$.
\item Sample $\pi_{ABC}(\theta|s_{obs})$ using the SMC-ABC routine proposed by \cite{toni2009approximate}, where $\hat{ \Sigma}(\theta_p)^{-1}$ is used as scaling matrix. We refer the reader to \cite{toni2009approximate} for details about this algorithm, but  point out that this is a sequential scheme where the tolerance $h$ is reduced at each step and that we terminated the algorithm when the acceptance ratio of the most recent iteration was below $1 \%$.
\end{enumerate} 
\end{itemize} 
We repeated the whole experiment 7 times and the results are illustrated in Figure \ref{fig: tol_vs_par}. Here the $x$-axis represents the value of $\log{(r)}$ at which the scaling matrix was estimated, while the $y$-axis represents the lowest tolerance $h$ achieved before the termination of the SMC-ABC algorithm. This plot shows how crucial is the choice of scaling matrix in situations where $ \Sigma( \theta )$ varies widely with $ \theta$: if the scaling matrix is not adequate the tolerance cannot be reduced enough. In an applied ecological setting, where the true parameters are unknown and the model of interest is more complex than the one used here, this means that a practitioner might struggle to find either a reasonable guess for the scaling matrix or a set of summary statistics whose covariance is not strongly dependent on $ \theta$. 

Another choice that has to be made, in order to use tolerance-based ABC procedures, is the selection of $h$. The tolerance can be a small scalar constant as in the MCMC-ABC algorithm of \cite{marjoram2003markov}, or it can be a vector of decreasing tolerances as in the SMC-ABC algorithm of \cite{toni2009approximate}. In order to obtain a better approximation to $\pi( \theta | s_{obs})$, $h$ should be chosen to be as small as possible, but the acceptance probability will decrease with the tolerance. A common choice is to select a tolerance that allows a predetermined acceptance ratio to be achieved, but in some cases this strategy can lead to invalid results, as detailed in \cite{silk2013optimizing}. 

The regression adjustment of \cite{beaumont2002approximate} \index{ABC!regression adjustment} can be used to mitigate the discrepancy between the observed and the simulated statistics, which is proportional to the tolerance $h$. However, the result of this correction is generally still dependent on $h$, which controls the bias-variance trade-off of the regression \citep{beaumont2002approximate}. Hence, using this procedure does not necessarily lead to higher accuracy in parameter estimation. For example, \cite{Fearnhead2012} obtained worse results with the regression correction than from the raw ABC output, using the Ricker model and the same summary statistics considered here.

SL is not afflicted by the difficulties just described, because it is tolerance-free and the summary statistics are scaled automatically and dynamically by the empirical covariance matrix $\hat{ \Sigma}(\theta)$. Obviously this robustness comes at a cost: a single point-wise synthetic likelihood estimate requires a number of simulations sufficient to estimate the covariance matrix. In addition, even though for many commonly used statistics the Central Limit Theorem (CLT) \index{Central Limit Theorem} assures asymptotic normality, in small samples the normal approximation might be crude, while in some contexts it might be difficult to devise asymptotically normal statistics. 

\begin{figure} 
\centering
\includegraphics[scale = 0.4, angle=-90]{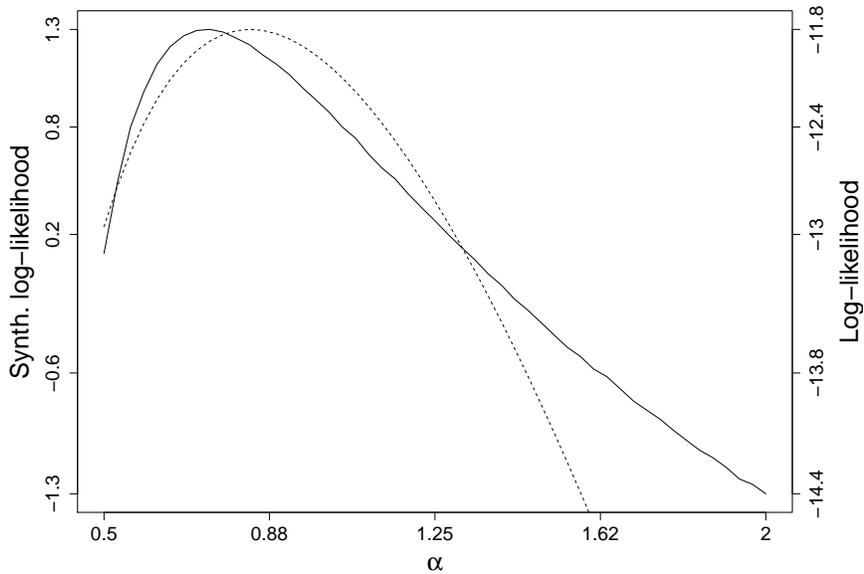}
\caption{Synthetic log-likelihood function (black line) vs true log-likelihood function (broken line) for a $\text{Exp}(\alpha = 1)$ distribution.}
\label{fig:exp_llk}
\end{figure}

As a simple example of the former problem, let us consider a sample of size $N$ from an exponential distribution with rate $\alpha$. Here the Maximum Likelihood (ML) estimator of $\alpha$ is given by the reciprocal of the sample average \index{SL!normal approximation}
$$
s = \frac{1}{\bar{x}} = \bigg ( \frac{\sum_{i = 1}^N x_i}{N} \bigg)^{-1}.
$$
Given that $s$ is a sufficient statistic for $\alpha$, the likelihood function can be factorized as follows
$$
p( x | \alpha) = h( x)f(s, \alpha) \propto f(s,\alpha),
$$
hence the likelihood is proportional to a function of only $s$ and $\alpha$. By the CLT, the distribution of $s$ is asymptotically normal, but we want to verify how well we can approximate the likelihood using SL when $N = 10$. Figure \ref{fig:exp_llk} shows the log-likelihood (dashed) and the estimated synthetic log-likelihood (black) for $\alpha \in [0.5, 2]$. The true value of $\alpha$ is 1. With such a small sample size the distribution of the simulated statistic is far from normal, and in fact the synthetic log-likelihood is quite off target. In cases such as this, where the number of summary statistics is low, it is straightforward to use transformations to improve to normality assumption, as proposed by \cite{wood2010statistical}. However, in an higher dimensional setting approximate multivariate normality might be difficult to assess or improve. More importantly, achieving multivariate normality for a certain set of parameters does not assure that this approximation will hold elsewhere in the parameter space. 

\section{Example: a chaotic prey-predator model} \label{sec:voles}

\index{model!prey-predator}

\index{dynamics!chaotic}

In order to illustrate the performance of SLMH and PMMH, we consider a modified version of the prey-predator model proposed by \cite{turchin2000}, which has been used to describe the dynamics of Fennoscandian voles (Microtus and Clethrionomys). \index{voles} More specifically, the model was an attempt at explaining the shift in voles abundance dynamics from low-amplitude oscillations in central Europe and southern Fennoscandia to high-amplitude fluctuations in the north. One of the possible drivers of this shift is the absence of generalist predators in the north, where voles are hunted primarily by weasels (Mustela nivalis) \citep{turchin2000}. According to this hypothesis, the lack of the stabilizing effect of generalist predators is the main factor determining the observed instability of voles abundances in the north.

The predator-prey dynamics are given by the following system of differential equations \citep{turchin2000}
$$
\frac{dN}{dt} = r(1 - e \sin{2 \pi t})N - \frac{r}{K}N^2 - \frac{GN^2}{N^2+H^2} - \frac{CNP}{N+D} + \frac{N}{K} \frac{dw}{dt},
$$
\begin{equation} \label{eq:volesModel}
\frac{dP}{dt} = s(1 - e \sin{2 \pi t})P - sQ\frac{P^2}{N},
\end{equation}
where $dw(t_2) - dw(t_1) \sim N[0, \sigma^2 (t_2-t_1)]$, with $t_2 > t_1$, is a Brownian motion process with constant volatility $\sigma$. The model is formulated in continuous time, because voles do not reproduce in discrete generations \citep{turchin1997empirically}. Here $N$ and $P$ indicate voles and weasels abundances, respectively. In the absence of predators, voles abundance grows at a seasonal logistic rate. Parameters $r$ and $s$ represents the intrinsic population growth rates of voles and weasels, while $K$ is the carrying capacity of the former. These parameters are averaged over the seasonal component, which is modelled through a sine function with amplitude $e$ and period equal to one year, with peak growth achieved in the summer. Generalist predation is modelled through a type III functional response, \index{functional response} under which generalists progressively switch from alternative prey to hunting voles, as voles density increases. The maximal rate of mortality inflicted by generalists is $G$, while $H$ is the half saturation parameter. 

Predation by weasels follows a type II response, where $C$ is the maximal predation rate of individual weasels and $D$ is the half saturation prey density. No prey-switching behaviour occurs under this functional response, which is consistent with weasels being specialist predators. Weasel abundance grows at a seasonal logistic rate, where the carrying capacity depends on prey density. Parameter $Q$ specifies the number of voles needed to support and replace an individual weasel and it determines the ratio of prey to predator densities at equilibrium. 

Differently from \cite{turchin2000}, who include environmental stochasticity in the system by randomly perturbing all model parameters using Gaussian noise with pre-specified volatility, we choose to explicitly perturb the prey equation using a Brownian motion process and to include its volatility $\sigma$ in the vector of unknown parameters. 

Vole abundance is not observed directly, but a proxy is provided by trapping data. We assume that the number of trapped voles is Poisson distributed
$$
Y_t \sim \text{Pois}(\Phi N_t),
$$
where $t \in \{ 1, \dots, T \}$ is the set of discrete times when trapping took place. No such proxy is available for weasels density, hence predator abundance represents a completely hidden state.

Following \cite{turchin2000} model (\ref{eq:volesModel}) is not fitted directly to data, but it is rescaled to a dimensionless form first. \index{dimensionless form} In particular, if we define
$$
n = \frac{N}{K}, \;\;\; p = \frac{QP}{K}, \;\;\; d = \frac{D}{K}, \;\;\; a = \frac{C}{K}, \;\; 
g = \frac{G}{K}, \;\;\; h = \frac{H}{K} \;\;\; \text{and} \;\;\; \phi = \Phi K,
$$
the reduced system is given by
$$
\frac{dn}{dt} = r(1 - e \sin{2 \pi t})n - r n^2 - \frac{gn^2}{n^2+h^2} - \frac{anp}{n+d} + n \frac{dw}{dt},
$$

$$
\frac{dp}{dt} = s(1 - e \sin{2 \pi t})p - s\frac{p^2}{N},
$$

\begin{equation} \label{eq:volesRedux}
Y_t \sim \text{Pois}(\phi n_t).
\end{equation}
%(XXX actually Turchin says that $p = \frac{KP}{Q}$, but I think they have confused the two Ps).
%
While \cite{turchin2000} implicitly re-scaled the simulations from the model, in order to match their means with that of the observed data, we formally estimate the scaling parameter $\phi$.

\cite{turchin2000} fitted the model by using a method which they call non linear forecasting (NLF), which is an instance of simulated quasi-maximum likelihood (SQML) method \citep{smith1993estimating}. One of the drawbacks of their estimation procedure is that it does not take into account the fact that trapping data provides noisy estimates of voles density. Another issue is that their method could not be used to estimate parameters that affect the variance of conditional distributions $p(n_t|n_{t-1}, n_{t-2}, \dots)$, but not their mean \citep{turchin2000}.

\subsection{Description of data and priors}

\begin{figure}
\centering
\includegraphics[scale = 0.40]{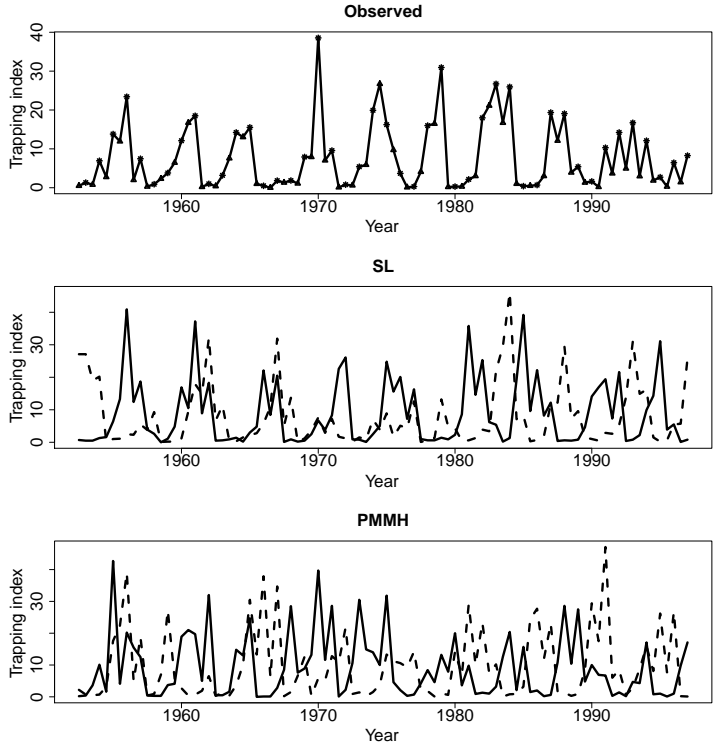}
\caption{Top: observed voles trapping index in Kilpisjarvi, between 1952 and 1997. Middle and bottom: two realization (solid and dashed) of model \ref{eq:volesModel}, using parameters equal to the posterior means given by SLMH and PMMH. }
\label{fig: voles_data}
\end{figure}

While \cite{turchin2000} consider several datasets, \index{data!Kilpisjarvi} \index{data!trapping} here we focus on the time series concerning voles abundance (mainly Clethrionomys rufocanus) in Kilpisjarvi, Finland. The data, shown in Figure \ref{fig: voles_data}, consists of 90 data points collected during the springs (mid-June) (triangles) and autumns (September) (stars) of each year, between 1952 and 1997. Each data point represents the number of voles trapped in a specific trapping season, divided by the number of hundred trap-nights used in that season. After 1980 the number of trap-nights was fixed to around 1000, but in earlier years this number is not available: it varied from a minimum of 500 to more than a thousand \citep{perry2000}. This correction for the sampling effort implies that, if the number of the trapped voles in each season is approximately Poisson distributed, the trapping index is not.

We have dealt with this problem by multiplying the data in Figure \ref{fig: voles_data} by 10 and by rounding each data-point to the nearest integer. This solution should give near-exact results for data collected after 1980, and a good approximation for all data-points representing a considerable population, thanks to the normal approximation to the Poisson distribution. 

A useful source of prior information is represented by \cite{turchin1997empirically}, where life history and data from short experiments were used to estimate the parameters of model (\ref{eq:volesRedux}). We report the prior distributions for each parameter in Table \ref{tab:volePrior}.
The expected values of the prior distributions has been chosen on the basis of the remarks of \cite{turchin1997empirically}, and we refer the reader to this reference for further details. The specific distributions and variabilities used for the priors have been chosen based on an attempt at quantifying the remarks of \cite{turchin1997empirically} regarding their confidence in their independently derived estimates. Admittedly, this process entails a certain degree of arbitrariness. No prior information was available for $\phi$ and $\sigma$, hence we have used improper uniform priors for both parameters.  

\begin{table}%1
\begin{center}
%\noautomaticrules
\begin{tabular}{cc}
Parameter    & Prior distribution \\
    $r$ & $N(\mu = 5, \sigma = 1)$  \\ 
    $e$ & $N(\mu = 1, \sigma = 1)$ \\ 
    $g$ & $\text{Exp}(\lambda = 7)$ \\ 
    $h$ & $\text{Gamma}(\kappa = 4, \theta = 40)$ \\ 
    $a$ & $N(\mu = 15, \sigma = 15)$ \\ 
    $d$ & $N(\mu = 0.04, \sigma = 0.04)$ \\ 
    $s$ & $N(\mu = 1.25, \sigma = 0.5)$ \\
    $\sigma$ & $\text{Unif}(0.5, \infty )$ \\ 
    $\phi$ & $\text{Unif}(0, \infty )$ \\ 
\end{tabular}
\caption{Priors used for the voles-weasels model.}%
\label{tab:volePrior}
\end{center}
\end{table}

For SL we used the following set of 17 summary statistics:
\begin{itemize}
\item autocovariances of $n_1,\dots,n_T$ up to lag 5;
\item mean population $\bar{n}$;
\item difference between mean and median population $\bar{n} - \tilde{n}$;
\item coefficients $\beta_1, \dots, \beta_5$ of the regression $n_{t+1}=\beta_1n_t + \beta_2n_t^2 + \beta_3n_{t-6} + \beta_4n_{t-6}^2 + \beta_5n_{t-6}^3 + z_t$;
\item coefficients of a cubic regression of the ordered differences $n_t-n_{t-1}$ on their observed values.
\item number of turning points, $\# n$.
\end{itemize}
This choice of statistics deserves some comments. Notice that, under suitable assumptions, all the above statistics are asymptotically normal as $T \rightarrow \infty$, due to the CLT. This provides some asymptotic justification to the Gaussian approximation used by SL. The autocovariances and the coefficient of the polynomial autoregressive model were meant to capture the dynamics of prey abundance on short ($\beta_{1, 2}$) and long ($\beta_{3,4,5}$) term basis. The degrees of the polynomials were choosed visually, by plotting $n_t$ against $n_{t-1}$ and $n_{t-6}$. Intermediate lags, such as $n_{t-3}$, were excluded, because they would have lead to very strong correlations between the regression coefficients. The marginal distribution of $n_t$ is summarized by $\bar{n}$ and $\bar{n} - \tilde{n}$, while the cubic regression coefficients aim at capturing the marginal structure of $n_t-n_{t-1}$. The number of turning points was introduced with the intention of capturing the volatility $\sigma^2$. This is because increasing $\sigma^2$ generally leads $\# n$ closer to $1/2$, which is typical of random walk behaviour.

\subsection{Comparison using simulated data}

\index{SL!vs PMMH}

In order to verify the accuracy of SLMH and PMMH for this prey-predator model, we have simulated 24 datasets of length $T = 90$, using parameters values $r = 4.5$, $e = 0.8$, $g = 0.2$, $h = 0.15$, $a = 8$, $d = 0.06$, $s = 1$, $\sigma = 1.5$ and $\phi = 100$. We have then estimated the parameters with both methods, using $2.5\times10^4$ MCMC iteration, the first $5\times10^3$ of which were discarded as burn-in period, and $10^3$ simulation from the model at each step. All the chains were initialized at the same parameter values. The resulting root mean squared errors (RMSEs) and variance-to-squared-bias ratios are reported in Table \ref{tab:voleMse}. While the RMSEs are quite similar for most parameters, the Table suggests that PMMH gives more accurate estimates for the scaling parameter $\phi$ and possibly for the generalist predation rate $g$. Indeed, SLMH estimates of $\phi$ are biased downward and are around ten times more variable than the estimates obtained with PMMH. In the case of $g$ the significance of the t-test should not be over-interpreted, given that it is attributable to PMMH achieving almost zero error on a single run.  

\begin{table}%1
\begin{center}
%\noautomaticrules
\begin{tabular}{ccccc}
Parameter & RMSE SLMH & RMSE PMMH & P-value & Best  \\
r & 0.33(3.3) & 0.25(9.9) & 0.49 & PMMH \\ 
  e & 0.19(0.1) & 0.2(0.1) & 0.78 & SLMH \\ 
  g & 0.09(0.2) & 0.08(0.5) & 0.05 & PMMH \\ 
  h & 0.04(0.2) & 0.03(0.4) & 0.15 & PMMH \\ 
  a & 2.12(1.3) & 1.97(1) & 0.48 & PMMH \\ 
  d & 0.02(0.5) & 0.02(0.6) & 0.57 & SLMH \\ 
  s & 0.07(18.6) & 0.08(10.9) & 0.22 & SLMH \\ 
$\sigma$ & 1.97(2.5) & 0.71(2.1) & 0.36 & PMMH \\ 
$\phi$ & 16.04(3.9) & 4.85(7.4) & $< 0.001$ & PMMH \\ 
\end{tabular}
\caption{RMSEs and variance-to-squared-bias ratios (in brackets) for SLMH and PMMH. P-values for differences in log-squared errors have been calculated using t-tests.}%
\label{tab:voleMse}
\end{center}
\end{table}

From a computational point of view, the two algorithms performed similarly. In particular, on a single 2.50GHz Intel i7-4710MQ CPU, point-wise estimates of $p(y_{obs}|\theta)$ or $p_{SL}(y_{obs}|\theta)$ cost around 1.55 and 1.35 seconds, when $10^3$ particles or simulated statistics are used. This time difference is marginal, and probably highly dependent on implementation details. However, it is worth pointing out that it is much easier to parallelize the computation of $\hat{p}_{SL}(s_{obs}|\theta)$ than that of $\hat{p}(y_{obs}|\theta)$. This is because of SIR's resampling step, which breaks the parallelisms at each time-step $t$ (see Box \ref{ABCreject}). For a review of parallelization strategies for the resampling step, see \cite{li2015resampling}. A possibly simpler solution is to compute several estimates $\hat{p}_1(y_{obs}|\theta)$, \dots, $\hat{p}_C(y_{obs}|\theta)$ in parallel, by running SIR with a fraction of the total number of particles $M$ on each of the $C$ cores, and then average them at each PMMH step to obtain a single estimate of $p(y_{obs}|\theta)$.

\subsection{Results from the Kilpisjarvi dataset}

\begin{figure}
\centering
\includegraphics[scale = 0.35]{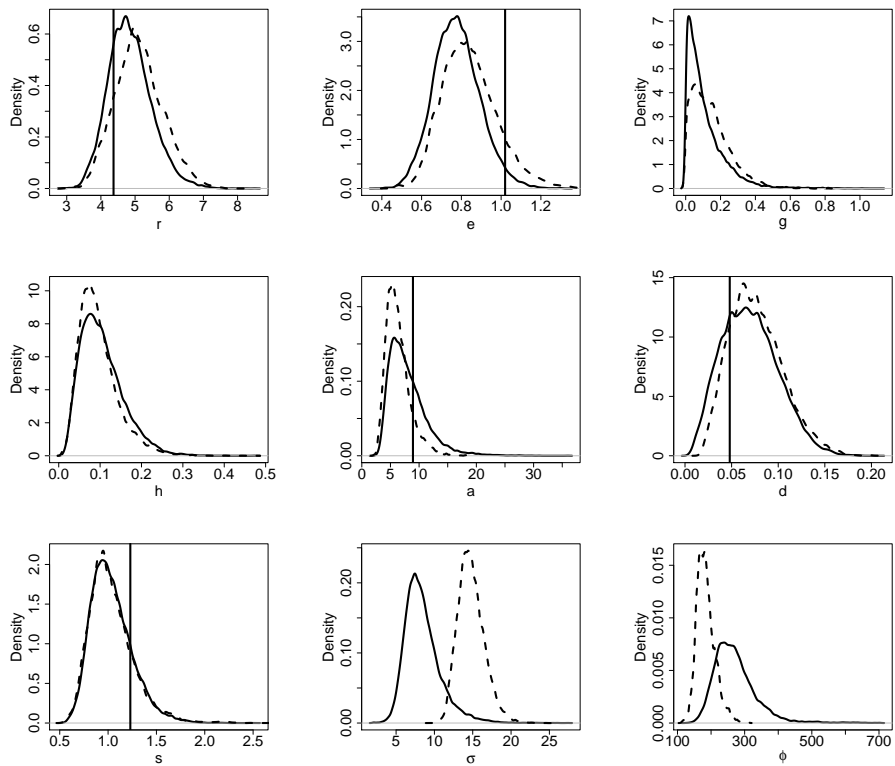}
\caption{Marginal posterior densities for voles model using SLMH (black) and PMMH (broke).
The vertical lines correspond to estimates reported by \protect \cite{turchin2000}, obtained using NLF (available only for 5 parameters).}
\label{fig: volesPost}
\end{figure}

We fitted the Kilpisjarvi dataset using $1.5 \times 10^5$ MCMC iteration, of which the first $10^4$ were discarded as burn-in period. At each step we used $10^3$ simulations from the model (SLMH) or particles (PMMH). The resulting posterior means are reported in Table \ref{tab:voleEstim}, while the marginal posterior densities of the parameters as shown in Figure \ref{fig: volesPost}. 

SLMH and PMMH give similar estimates for most parameters, with substantial differences only for $\sigma$ and $\phi$. Indeed, PMMH's estimate of the former parameter is much higher than that obtained using SL. Interestingly, \cite{fasiolo2014statistical} encountered a similar pattern when fitting the blowfly model of \cite{wood2010statistical} to Nicholson's experimental datasets \citep{nichol54,nichol57}. In that context, the process noise estimates were much higher under PMMH than under SL, on all datasets. This biased PMMH's estimates of the remaining parameters towards stability, particularly on two of the datasets. As we will show later in this section, this stabilizing effect of high process noise estimates on the dynamics is less noticeable here.
  
Figure \ref{fig: voles_data} compares the observed data with trajectories simulated from model (\ref{eq:volesModel}), using parameters equal to the posterior means given by SLMH and PMMH. Both methods seem to produce dynamics that are qualitatively similar to the observed ones, with the paths simulated using PMMH's estimates being slightly more irregular, which is attributable to the higher process noise estimate. 

\begin{table}%1
\begin{center}
%\noautomaticrules
\begin{tabular}{cccccc}
 & $ r$ & $ e$ & $ g$ & $ h$ & $ a$   \\
  SLMH & 4.85(0.63) & 0.78(0.12) & 0.11(0.11) & 0.1(0.05) & 8.0(3.3)  \\ 
  PMMH & 5.11(0.7) & 0.84(0.14) & 0.14(0.11) & 0.1(0.05) & 6.3(2.1)  \\ 
 & $ d$ & $ s$ & $ \sigma$ & $ \phi$ & \\
  SLMH & 0.07(0.03) & 1.04(0.21) & 8.4(2.3) & 270.5(63.5) & \\ 
  PMMH & 0.08(0.03) & 1.04(0.23) & 14.8(1.7) & 184.2(26.9) & \\ 
\end{tabular}
\caption{Estimated posterior means (standard deviations) for model \ref{eq:volesModel}.}%
\label{tab:voleEstim}
\end{center}
\end{table}

\index{SL!diagnostics}
\index{summary statistics!normality}

\begin{figure}
\centering
\includegraphics[scale = 0.5]{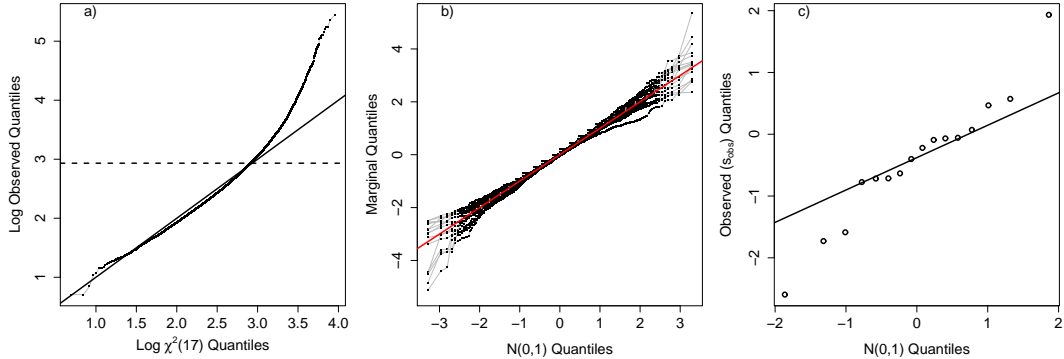}
\caption{Normality plots for the simulated summary statistics. See main text for details.}
\label{fig:qqNorm}
\end{figure}

Beside comparing observed and simulated trajectories, in the context of SL it is also advisable to check whether the summary statistics are indeed approximately normally distributed. In particular, it is important to verify whether this assumption holds within the highest posterior density region. For this reason, we simulated $M = 10^4$ summary statistics, $S^{1:M}=\{S^1, \dots, S^M\}$, from the model, using parameters equal to the estimated posterior mean. Then we used the methods of \cite{krzanowski1988principles} to produce the normality plots shown in Figure \ref{fig:qqNorm}. In particular, we transformed $S^{1:M}$ to a variable that should be $\chi^2(17)$, under normality of $S$. Figure \ref{fig:qqNorm}a compares observed and theoretical log-quantiles. Departures on the right end of the plot indicate that the normal approximation is poor in the tails. In this case this is not much of problem, as the dashed line, which indicates the Mahalonobis distance between $s_{obs}$ and the sample mean of $S^{1:M}$, falls within the region where the normal approximation is adequate. Figure \ref{fig:qqNorm}b shows marginal normal q-q plots for the simulated statistics. Marginal normality seems to hold reasonably well for most statistics, with the exception of $\beta_5$, whose distribution is skewed to the left. An analogous qq-plot for normalized observed statistics $s_{obs}$ is shown in Figure \ref{fig:qqNorm}c. This plot is suggestive of departures from normality, but this approach does not have much power, unless the dimension of $s_{obs}$ is fairly large.

\begin{figure}
\centering
\includegraphics[scale = 0.40]{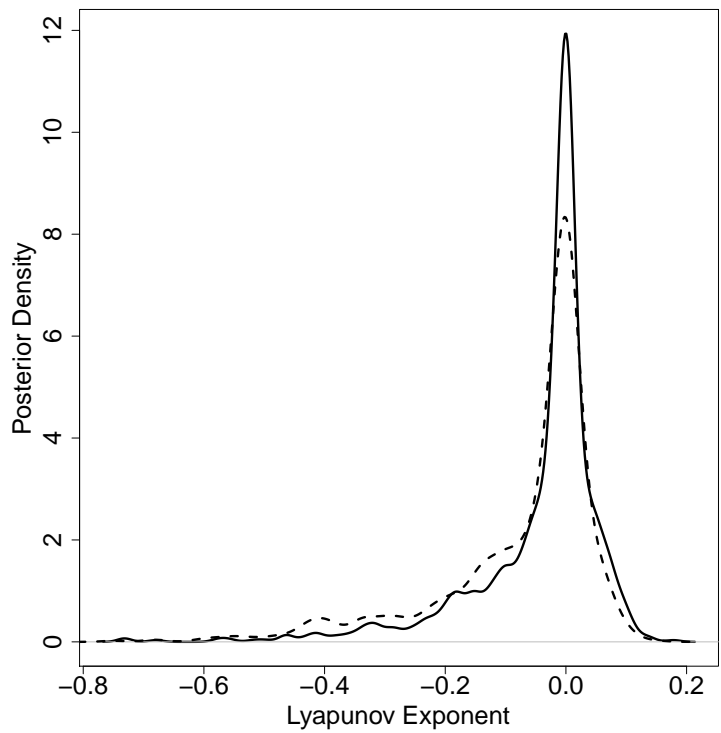}
\caption{Approximate posterior densities of Lyapunov exponents for SLMH (black) and PMMH (broken).}
\label{fig:lyapunov}
\end{figure}

One of the main scientific questions model ($\ref{eq:volesModel}$) was meant to address was whether the observed dynamic in voles densities can be classified as chaotic. To answer this question we have randomly sampled $10^3$ parameters sets from posteriors samples obtained by SLMH and PMMH. We have then used each parameters set to simulate a trajectory from the deterministic skeleton of model ($\ref{eq:volesModel}$) for $10^5$ months, which were discarded in order to let the system leave the transient, and used additional $10^4$ months of simulation to estimate the maximal Lyapunov exponent as in \cite{wolf1985}. By doing this, we obtained the two approximate posterior densities of the Lyapunov exponent shown in Figure \ref{fig:lyapunov}. Notice that the posterior produced by PMMH is slightly more skewed to the left relatively to that obtained with SL, which suggests that the system dynamics are estimated to be more stable under the former methods. Together with the high estimate of $\sigma^2$, this confirms the tendency of PMMH to inflate the noise and to bias the estimated dynamics toward stability. While this effect was very pronounced under the blowfly model studied by \cite{fasiolo2014statistical}, in this case it is very mild. Indeed, the median Lyapunov exponent is equal to $-6\times 10^{-4}$ for SLMH and $-0.015$ for PMMH. These estimates are very close to each other and to the one ($-0.02$) reported by \cite{turchin2000} for this dataset, and provide more model-based evidence supporting the hypothesis that this system lives on the edge of chaos.

\index{Lyapunov exponent}

\section{Discussion} \label{ref:discussion}

The example presented in this work gives the flavour of what can be accomplished using SL or particle filters, in the context of ecological SSMs. Both approaches provided a sample from the parameters' posterior distribution, which is the result of a full Bayesian analysis that incorporates both prior and likelihood-based information. While Table \ref{tab:voleMse} suggests that SL might have lost information regarding some of the parameters, in Section 1.2.3 we point out that the estimates provided by PMMH might be slightly biased towards stability. In essence, PMMH estimates the process noise $\sigma^2$ to be quite high, which leads PMMH to explain the observed dynamics using noise, rather than by moving the system away from stability, using other dynamically important parameters. When dealing with highly non-linear models, it is worth being aware of this tendency, because it can be lead to system dynamics being classified as stable even when they are not, as shown by \cite{fasiolo2014statistical} using the blowfly model of \cite{wood2010statistical}. Fortunately, SLMH and PMMH strongly agree in classifying the dynamics of the prey-predator system considered in this work as near-chaotic, hence either approach could have been used to answer the main scientific question underlying model (\ref{eq:volesModel}).

From the point of view of applied ecologists, ease of use and automation are arguably as important as statistical and computational efficiency. In Section \ref{sec:voles} we have shown that the choice of scaling matrix can be very important for ABC methods. Selecting this parameter correctly can be particularly difficult when little or no prior knowledge about model parameters is available. From this point of view, SL is at an advantage with respect to other ABC methods because, once the summary statistics have been selected, there is very little tuning to do. Obviously SL pays for this tuning-free property with a normality assumptions, which might result in lower accuracy.

The summary statistics selection process, which SL cannot escape, can be the most time consuming and arbitrary step of the inferential process. \index{summary statistics!selection} In the example presented in Section \ref{sec:voles} we obtained good results, in terms of parameter accuracy, by using the statistics of \cite{wood2010statistical} with some modifications. In our experience this is the exception, rather than the rule. In fact, even though \cite{blum2013comparative} offer several systematic approaches for statistics selection, studying the model output by visualizing characteristics such as empirical transition densities, periodicity and dependencies between states is still indispensable for most models of reasonable complexity. 

ABC methods have become popular tools for dealing with complex phylogeographic \citep{hickerson2010phylogeography}, phylogenetic \citep{rabosky2009heritability} and individual based \citep{hartig2014technical} models, but they do not seem to have been equally successful for dynamical SSMs of ecological interest. The main reasons for this might be that particle filters represent an obvious alternative, and that at the moment it is not clear whether ABC methods can outperform them along any dimension of the inferential process. In fact, particle filters have the important advantage of using the full data, $y_{obs}$, thus avoiding both the information loss and the issue of choosing the summary statistics. On the other hand, this use of all the data makes filtering more susceptible to model mis-specification problems, in which failures to capture the data generating mechanism exactly can have a substantial negative impact on inference. \index{model!mis-specification}

The robustness properties of methods based on summary or ``intermediate'' statistics, in particularly the protection they can offer against model mis-specification and outliers, has beed widely recognized and exploited in econometrics, but it seems to have attracted less attention in the wider statistical community \citep{jiang2004indirect}. Hence, it would be interesting to verify whether ABC methods share any of the robustness properties of more traditional statistics-based approaches. If this turns out to be the case, one possibility is that ABC methods will be used in support of more accurate, but possibly less robust, methods based on the full likelihood, such as particle filters. This was suggested by \cite{fasiolo2014statistical}, in the context of highly non-linear ecological and epidemiological models, and by \cite{owen2014scalable}, who propose a hybrid procedure where an ABC sampler is used in support of a PMMH algorithm. While both works have suggested that ABC methods are more robust than particle filters to bad initializations, the first one has also found that they are less affected by outliers and that they can provide reliable parameter estimates when dealing with highly non-linear models characterized by extremely multimodal full likelihoods. \index{summary statistics!robustness}

Although the use of summary statistics wastes information and requires an often time-consuming statistics selection process, ABC methods have some features that are very appealing from a practical perspective. In fact, they are purely simulation-based or ``plug-and-play'' \citep{bhadra2011malaria}, because they only require  simulation of data from the model and transformation to summary statistics. \index{plug and play} This property makes these methods general-purpose, because they can be used to fit any model for which a simulator is available, with little or no assumptions required. Hence ABC methods can potentially accelerate the model development process: once the summary statistics have been chosen, testing new model versions requires only updating the simulator. In addition, this generality allows practitioners to explore models that violate the assumptions necessary for particle filters to work, such as Markovian dynamics or the tractability of the observational density $p(y_t|x_t)$. 

Similar practical considerations hold also in regard to the programming effort necessary to implement each method. For models of moderate complexity, no ABC or particle filtering method can be entirely implemented in a traditional interpreted language (such as \emph{R}). In fact, any of these methods requires at least part of the code to be written in a compiled language (such as \emph{C/C++}). In the case of ABC methods this is often simple to do, because the largest share of the computational time is spent simulating data and transforming it to summary statistics, so it is often sufficient to write only these steps in a compiled language. On the other hand particle filters generally do not simulate whole datasets, but work in sequential steps, so it is difficult to isolate the parts of these algorithms that have to be implemented efficiently. This means that it might be necessary to write these procedures entirely in a compiled language, which slows down the model development and evaluation process. 

For these reasons, software tools providing frameworks and algorithms for doing inference for SSMs are very useful to statistical ecologists. One such example is the \emph{pomp} \emph{R} package \citep{pompPackage}, which we used to set up the model described in Section \ref{sec:voles}. This package focuses mainly on tools based on particle filtering, but it offers also several approximate approaches, and it can greatly reduce the programming effort, if the model of interest fits the framework provided by the package. While statistical suites are available for tolerance-based ABC methods and for SL, such as the \emph{EasyABC} \citep{easyabcPackage} and the \emph{synlik} \citep{synlikPackage} \emph{R} packages, \index{SL!software} these do not focus on SSMs in particular, thus reflecting the wide range of application of the underlying statistical methodologies.

In conclusion, ABC methods offer an approach to intractable ecological models that forgo information in exchange for generality and, possibly, robustness. While this trade-off has shown to be fruitful in many branches of ecological modelling \citep{hartig2011statistical}, particularly when the model is not intended to reproduce the data exactly, future work will determine whether ABC methods will play a major role in the context of SSMs, possibly alongside less approximate approaches.

\section*{Acknowledgments}

This work was performed under partial support of the EPSRC grant EP/I000917/1 and EP/K005251/1.

\bibliographystyle{chicago}
\bibliography{biblio}

\end{document}